\documentstyle[twoside,fleqn,espcrc2]{article}

\newcommand{\beq}{\begin{eqnarray}}
\newcommand{\eeq}{\end{eqnarray}}
\newcommand{\nn}{\nonumber}

\newcommand{\dd}{\partial}

\newcommand{\conss}
{\sbra{\prod_x \delta_{\partial'_{\mu}\tilde{\sigma}_{\mu\nu}(x) =
0}}}

\newcommand{\sbra}[1] { \left( #1 \right)}    
\newcommand{\mbra}[1] { \left\{ #1 \right\}}  
   
\newcommand{\zbra}[1] { \left| #1 \right|}   
  
\newcommand{\kbra}[1] { \left< #1 \right>}

\newcommand{\AmS}{{\protect\the\textfont2
  A\kern-.1667em\lower.5ex\hbox{M}\kern-.125emS}}

\title{
On the perfect lattice actions 
of abelian-projected SU(2) QCD
\thanks{presented by S. Kato}}
\author{
S. Fujimoto, 
\hspace{2mm}
\thanks{ E-mail address:kato@hep.s.kanazawa-u.ac.jp }
S. Kato, 
\hspace{2mm}
M. Murata
\hspace{2mm}
and \hspace{2mm}
T. Suzuki \\ 
\vspace{3.mm}
Department of Physics, Kanazawa University Kanazawa 920-1192, Japan
}

\begin{document}

\begin{abstract}
We study the perfect lattice 
actions
 of abelian-projected 
SU(2) gluodynamics. 
Using the BKT and duality transformations on the 
lattice, an
effective string model is derived 
from 
the
{\bf direction-dependent} quadratic monopole 
action, 
obtained numerically from SU(2) gluodynamics in maximally
abelian gauge.
The string tension and the restoration of continuum rotational 
invariance are investigated using strong coupling expansion of 
lattice string model {\bf analytically}.
We also found that the block spin transformation can be performed
{\bf analytically} for the quadratic monopole action.
\end{abstract}

\maketitle

\input epsf

{\bf \hspace{-0.35cm}1.$\ $INTRODUCTION}

The infrared effective theory of QCD is important for 
the analytical understanding of  hadron physics. Abelian monopoles 
which appear after
abelian projection of QCD \cite{'thooft} seem to be relevant
dynamical degrees of freedom for infrared region \cite{domi}. 
Shiba and Suzuki \cite{shiba_suzuki} derived the monopole 
action from vacuum configurations obtained in Monte-Carlo simulations
extending the method developed by Swendsen.

We studied the renormalized monopole action performing 
block spin transformations up to $n=8$ numerically, 
and saw that scaling for fixed $b$ looks good\cite{nakam}.
If the effective action obtained here is very near to 
the
perfect
action, the physical quantities from it should reproduce the 
continuum rotational symmetry, although the action is formulated 
on the lattice. In order to show restoration of 
rotational invariance,
the direction-dependence of the renormalized monopole action 
is very
important. Practically, it is difficult to evaluate 
the
string tension numerically using monopole action 
for infrared region, since
Wilson loop operators follow 
a simple exponential curve and 
they become too small
within the statistical noise. So we 
try to
evaluate 
the Wilson loops
using 
the 
string model corresponding to 
the
renormalized monopole action.
%

\vspace{0.4cm}
{\bf \hspace{-0.35cm}2.$\ $STRING REPRESENTATION FROM \\
\hspace*{0.4cm}MONOPOLE ACTION}

 We derive here the lattice string model using BKT(Berezinskii
-Kosterlitz-Thouless) transformation.

Let us start from the following direction-dependent monopole
partition function;  
\beq
 {\cal Z} =\hspace{-0.4cm} \sum_{k_{\mu}(x)=-\infty \atop 
                   {(\partial_{\mu}'k_{\mu}(x)=0)}}
                    ^{\infty}\hspace{-0.4cm} 
           \exp\mbra{-\sum_{x,y}\sum_{\mu=1}^D k_{\mu}(x)
            {\cal D}(x,y;\hat{\mu})k_{\mu}(y)
}, \nn
\eeq
where $D=4$ is space-time dimensionality and closed monopole 
currents $k_{\mu}(x)$  are defined on the dual lattice.
Operator ${\cal D}(x,y;\mu)$ is composed of direction-independent 
part ${\cal D}_{1}$ and -dependent part ${\cal D}_{2}$;
\beq
&& {\cal D}(x,y;\hat{\mu}) \equiv {\cal D}_{1}(x-y)
   +{\cal D}_{2}(x-y;\hat{\mu}),  \nn \\
&&  {\cal D}_{2}(x,y;\hat{\mu}) = g_1(b)\cdot \delta_{x,y} 
     + g_2(b) \cdot \delta_{x+\hat{\mu},y} \nn \\
&&     + g_3(b) \cdot \sum_{\gamma \ne \mu}\delta_{x+\hat{\gamma},y} 
     + \cdots , \nn
\eeq
where ${\cal D}_{1} >> \zbra{{\cal D}_{2}}$ and 
$g_2(b) \ne g_3(b)$ etc. .

$b(\beta,n)=na\sbra{\beta}
=\sqrt{\tilde{\kappa}(\beta,n)/\kappa_{phys}}$
is the physical length in unit of the physical string tension
\(\kappa_{phys}\). The dimensionless string tension 
\(\tilde{\kappa}\) is determined by the lattice Monte-Carlo 
simulation, \(a(\beta)\) is the lattice spacing and \(n\) is the
number of blocking steps.
The couplings of the above monopole action are determined 
using the 
extended Swendsen method([3-4])
%
We have found that 
four and six point interactions are very small 
for low-energy region of QCD\cite{nakam}. Hence we consider above
only quadratic interactions for monopole currents for simplicity.

Using the transformation suggested in Ref.~\cite{PWZ},
this type of monopole action can be transformed into the 
string model;
\beq
&&\!\!\!\!\!\!\!\!\!\!\!
{\cal Z} = \mbox{const.}
\hspace*{-0.5cm}\sum_{\tilde{\sigma}_{\mu\nu}(x)=-\infty}
    ^{\infty}\hspace{-0.2cm} \conss \exp\left\{ {\cal S}_{STR}\right\} \nn \\
&&\!\!\!\!\!\!\!\!\!\!\!
{\cal S}_{STR} = -\pi^2 \!\sum_{x,y}\sum_{\mu < \nu}
            \!\tilde{\sigma}_{\mu\nu}(x)
            \!\!\sbra{\frac{1}{\Delta {\cal D}_{1}}}\!\!(x-y)
            \tilde{\sigma}_{\mu\nu}(y) \nn \\ \nn \\  
&&\!\!\!\!\!\!\!\!\!\!\!
\hspace*{0.7cm}- \frac{\pi^2}{4}
    \sum_{x,y}\sum_{\mu \ne \nu}   
     \epsilon_{\mu\nu\xi\eta}
     \epsilon_{\mu\alpha\gamma\delta}
     \tilde{\sigma}_{\xi\eta}(x-\hat{\xi}-\hat{\eta})
           \nn \\ 
&&\!\!\!\!\!\!\!\!\!\!\!
\quad\quad  \times
     \sbra{\frac{{\cal D}_{2}}{(\Delta {\cal D}_{1})^2}}
     (x-y;\hat{\mu})
     \dd'_{\alpha}\dd_{\nu}
     \tilde{\sigma}_{\gamma\delta}(y-\hat{\gamma}-\hat{\delta})
      \nn \\ \nn \\
&&\!\!\!\!\!\!\!\!\!\!\!
\hspace*{0.7cm} - (\mbox{higher order term}) \nn  
\eeq
where we used
\beq
{\cal D}^{-1} 
= {\cal D}_{1}^{-1} 
+ {\cal D}_{1}^{-1}{\cal D}_{2}{\cal D}_{1}^{-1}+ \cdots \nn
\eeq

The integer-valued plaquette field $\tilde{\sigma}_{\mu\nu}(x)$ 
which is defined on the original lattice represents the 
closed world surface formed by a color electric flux tube.
The leading part of this model comes from the direction-
independent part of the monopole action; the next-leading 
terms come from the contribution of 
the
direction-dependent part.

\vspace{0.4cm}
{\bf \hspace{-0.35cm}3.$\ $ROTATIONAL INVARIANCE}

 In order to check the restoration of continuum rotational 
symmetry, let us consider the $q$-$\bar{q}$ static potential 
at
 the points $(2,0,0)$ and $(1,1,0)$ of a three-dimensional 
time-slice, respectively. (The quark is attached on the origin $(0,0,0)$ 
and antiquarks are attached on $(2,0,0)$ and $(1,1,0)$, 
respectively.)

The static potential $V(x,y,z)$ are calculated from 
the Wilson loop 
operators: 
\beq
&&\!\!\!\!\!\!\!\!\!\!\!
V(2,0,0) = -\lim_{T\to\infty}\frac{1}{T}\log<W(2,0,0,T)> \nn \\
&&\!\!\!\!\!\!\!\!\!\!\!
V(1,1,0) = -\lim_{T\to\infty}\frac{1}{T}\log<W(1,1,0,T)>. \nn
\eeq
If the potential is purely linear and the continuum rotational
symmetry is restored, then the ratio $V(2,0,0)/V(1,1,0)$ should
become $\sqrt{2}$. 

The quantum average of Wilson loop operator in the string 
representation is written as follows.
\beq
&&\!\!\!\!\!\!\!\!\!\!\!
\kbra{W(C)} = \frac{1}{{{\cal Z}_{STR}}}
\hspace{-0.3cm}\sum_{\tilde{\sigma}_{\mu\nu}(x)=-\infty
      \atop {\left(\partial'_{\mu}\tilde{\sigma}_{\mu\nu}(x) =
      \tilde{J}_{\nu}(x)\right)}}^{\infty}
\hspace{-0.5cm}\exp\mbra{{\cal S}_{STR}[\tilde{\sigma}_{\mu\nu}(x)]} \nn
\eeq
Note that 
the
string field $\tilde{\sigma}_{\mu\nu}(z)$ are
forming open surfaces whose boundaries are 
the
Wilson loop.
We 
can
evaluate this quantity using {\bf strong coupling expansion}.

At $b=2.14$ ( $\cong$ 0.96 fm ), as a preliminary result, the string 
tension from $V(2,0,0)$ become $1.5$ in unit of 
$\kappa_{phys}$
and the ratio $V(2,0,0)/V(1,1,0) = 1.07$. 
The 
quadratic part of the
renormalized direction-dependent monopole action 
does not seem
to reproduce the correct string tension
and the continuum rotational invariance.
We probably need 
(1) to consider 4 and 6 point interactions,
(2) more steps of blocking on larger
lattice volume for larger $\beta$ and 
(3) more complicated form of monopole action. 
%

\vspace{0.4cm}
{\bf \hspace{-0.35cm}4.$\ $ANALYTIC BLOCK SPIN TRANSFORMATION}

We found that the block spin transformation can be performed {\bf
analytically} for the quadratic monopole action.

When $b$ is large, the London limit of the abelian Higgs model
works good as an effective theory of QCD.
 For example, let us start from the following simple monopole 
partition function defined on the small $a(\beta)$-lattice:
\begin{eqnarray*}
&&\!\!\!\!\!\!\!\!\!\!\!
Z=\hspace{-0.3cm}\sum_{a^3k=-\infty \atop {\left(\partial_{\mu}'k_{\mu}(x)=0\right)}}
^{\infty}\, 
  \;\hspace{-0.3cm}\exp\left\{ -\sum_{s,s';\mu}\, k_\mu(s) D(s-s')
       k_\mu(s')\right\} \\
&&\!\!\!\!\!\!\!\!\!\!\!
 D(p)\!=\!\left(\!\alpha\!+\frac{\beta}{4\sum_{\rho}\sin^2(\frac{p_\rho}{2})} \right)
        \!\!\!\left(\!1\!+\!4\epsilon \,\!\!\sum_{\rho}\sin^2(\frac{p_\rho}{2}) \right)
\end{eqnarray*}
This action corresponds to the London limit of the abelian-Higgs
model, when $\epsilon = 0$.

Using the Poisson summation formula, the monopole action defined 
on the $b=n a(\beta)$ lattice is given by
\begin{eqnarray*}
&&\!\!\!\!\!\!\!\!\!\!\!
 e^{-S[K]}
  =\hspace{-0.7cm}\sum_{a^3k_\mu(\!as\!)=-\!\infty}^{\infty}\!\!\!\!
  \hspace{-0.2cm}\!\delta\left(\!\sum_{\mu}\!\partial'_{\mu}k_{\mu}(as) \right) 
  \!\delta\Bigg(
    b^3K_{\mu}(bs)\!-\!{\cal M}\!\Bigg)  \\
&&\!\!\!\!\!\!\!\!\!\!\!
\hspace{0.5cm}\exp\bigg\{\!-\!(a^4)^2\sum_{s,s';\mu} k_\mu(as) D(as-as') k_\mu(as')
    \bigg\}
\end{eqnarray*}
\vspace{-0.7cm}\\
where,
\vspace{-0.2cm}
\begin{eqnarray*}
&&\!\!\!\!\!\!\!\!\!\!\!
{\cal M}\equiv\!\!\!\sum_{i,j,l=0}^{n-1}\!\!a^3
k_\mu\Big(nas\!+(n\!-\!1\!)a\mu\!+\!ia\nu\!+\!ja\rho\!+\!la\sigma\!\Big),\\
&&\!\!\!\!\!\!\!\!\!\!\!
S[K]=\!\!\int_{-\pi /na}^{+\pi /na}\!\!\frac{d^4p}{(2\pi)^4}
         K_{\mu}(-p) \big[\triangle^{GF}_{\mu\nu}(p)\big]^{-1} K_{\nu}(p) \\
&&\!\!\!\!\!\!\!\!\!\!\!
\end{eqnarray*}
\vspace{-0.7cm}\\
and
\vspace{-0.2cm}
{\footnotesize
\begin{eqnarray*}
&&\!\!\!\!\!\!\!\!\!\!\!
\triangle_{\mu\nu}(p)=
\frac{1}{(\alpha-\beta\epsilon)}
\{\bar{\triangle}_{\mu\nu}(p,\epsilon^{-1})
-\bar{\triangle}_{\mu\nu}(p,\frac{\beta}{\alpha})\} \\
&&\!\!\!\!\!\!\!\!\!\!\!
\bar{\triangle}_{\mu\nu}(p,m^2) \equiv 
\frac{m^2}{n^6}\sum_{l=0}^{n-1}
\frac{\Pi_{\neg\mu}(p+2\pi l)\Pi_{\neg\nu}(p+2\pi l)}
{4\sum_\rho\sin^2(\frac{p_\rho a}{2} +\frac{\pi l_\rho}{n})+m^2}\\
&&\!\!\!\!\!\!\!\!\!\!\!
\times \biggl(
	 \delta_{\mu\nu}\hspace{-0.1cm}
         -\hspace{-0.05cm}\frac{\sin(p_\mu a +\frac{\pi l_\mu}{n})
                \sin(p_\nu a +\frac{\pi l_\nu}{n})}
	       {\sum_\rho\sin^2(p_\rho a +\frac{\pi l_\rho}{n})}
\biggr)e^{\frac{i}{2}(p_\mu-p_\nu)na} \\
&&\!\!\!\!\!\!\!\!\!\!\!
 \Pi_{\neg\mu}(p)\equiv
\prod_{i\ne \mu}\frac{\sin(p_i na/2)}{\sin(p_i a/2)}.
\end{eqnarray*}}
Using the above analytical block spin transformation, we can find
a lattice monopole action which reproduces the continuum 
rotational invariance with the correct string tension
if we take $n\to\infty$ and $a(\beta)\to 0$ for fixed 
$b=n\cdot a(\beta)$\cite{wiese}. 

Let us evaluate the string tension and the ratio of the 
potential from string model 
expression corresponding to $S[K]$ above.
The ratio $V(2,0,0)/V(1,1,0) = 1.24$ at $b=2.14$( $\cong$ 0.96 fm ).
\footnote[3]{The parameters $\alpha = 0.73$,$\beta = 0.73$ and 
$\epsilon = 0.01$ are determined so that the string tension becomes
unity in unit of physical one at $b=2.14$.}
It 
approaches considerablly to the value expected from
the continuum rotational invariance. The 
discrepancy may be due to truncation effect of the monopole action.

\begin{figure}
\epsfysize=70mm
 \vspace{-20pt}
 \begin{center}
 \leavevmode
\epsfbox{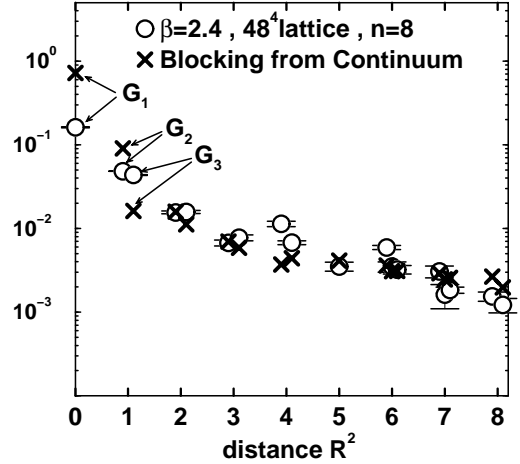}
 \end{center}
 \vspace{-40pt}
{\footnotesize
\caption{
The \(R^2\) dependence of the monopole action
from Swendsen's method and analytical block spin transformation
at $b=2.14$.
\label{Mono}
}}
\vspace{-15pt}
\end{figure}

\vspace{0.4cm}
{\bf \hspace{-0.35cm}5.$\ $CONCLUSIONS}

Let us compare the monopole action 
$S[k]$
in previous section 
with the present numerical data fixed by the Swendsen 
method. We see the self coupling $G_1$ and the
discrepancy between nearest-neighbor couplings($G_2$ and
$G_3$) are larger than those of the monopole action determined
numerically
as shown in
Figure 1. These behavior 
must
be 
important to reproduces the continuum rotational invariance.

Recently, the spectrum of glueball masses in non-supersymmetric
Yang-Mills theory based on a Maldacena's conjectured
duality between supergravity and large N gauge theories
are evaluated\cite{ooguri}.
Our string model also yields glueball mass spectrum analytically 
using the strong coupling expansion of the correlation functions 
of gauge invariant local operators. This is now
in progress.


\end{document}